\documentclass[useAMS,usenatbib,aas_macros]{mn2e}
\input{psfig}
\usepackage{aas_macros}
\usepackage{natbib}
\usepackage[dvips]{epsfig}
\usepackage{afterpage}
\usepackage[T1]{fontenc}
\usepackage{aecompl}

\title[A new period of activity in the core of NGC\,660]{A new period of activity in the core of NGC\,660}

\author[M.K. Argo et al]{Megan K. Argo$^{1}$, Ilse M. van Bemmel$^{2,3}$, Sam D. Connolly$^{4}$ and Robert J. Beswick$^{1}$\\ 
1. Jodrell Bank Centre for Astrophysics, School of Physics and Astronomy, The University of Manchester, Oxford Road, Manchester, M13 9PL, UK\\
2. Netherlands Institute for Radio Astronomy (ASTRON), Postbus 2, 7990 AA Dwingeloo, The Netherlands\\
3. Joint Institute for VLBI in Europe (JIVE), Postbus 2, 7990 AA Dwingeloo, The Netherlands\\
4. School of Physics and Astronomy, University of Southampton, Highfield, Southampton, SO17 1BJ, UK\\}

\def\farcs    {\hbox{$.\!\!^{\prime\prime}$}}
\def\kms      {\ifmmode {\rm km\,s}^{-1} \else km\,s$^{-1}$\fi}

\def\rasec    {\hbox{$.\!\!^{\rm s}$}}
\def\degr     {\hbox{$^\circ$}}

\begin{document}
\maketitle

\begin{abstract} 
{
The core of the nearby galaxy NGC\,660 has recently undergone a spectacular radio outburst; using a combination of archival radio and Chandra X-ray data, together with new observations, the nature of this event is investigated.  Radio observations made using e-MERLIN in mid-2013 show a new compact and extremely bright continuum source at the centre of the galaxy.  High angular resolution observations carried out with the European VLBI Network show an obvious jet-like feature to the north east and evidence of a weak extension to the west, possibly a counter-jet.  We also examine high angular resolution H{\sc i} spectra of these new sources, and the radio spectral energy distribution using the new wide-band capabilities of e-MERLIN.  We compare the properties of the new object with possible explanations, concluding that we are seeing a period of new AGN activity in the core of this polar ring galaxy.
}
\end{abstract}



\section{A polar ring galaxy}

The nearby galaxy NGC\,660 ($\sim$13\,Mpc; \citealt{2009ApJS..182..474S}) exhibits a polar ring morphology and a LINER-type nucleus (low-ionisation nuclear emission-line region; \citealt{2002A&A...392...53N}).
\cite{1992A&A...259L..65C} used observations of CO emission to infer a total H$_{2}$ mass in the main disk (radius 4\,kpc) of 3.7 $\times 10^{9}$\,M$_{\odot}$, and in the polar ring (radius 12\,kpc) of 10$^{9}$ M$_{\odot}$.  They construct a model to account for both the CO and H{\sc i} properties of the system, concluding that the ring is highly inclined and highly warped.
The stellar population in the ring is estimated to be a few billion years old \citep{1995AJ....109..942V}.

The large-scale H{\sc i} and OH properties were probed by \cite{1992ApJ...401..508B} who saw a large velocity gradient from the disk in absorption, as well as absorption from the warped outer disk at the systemic velocity.  The same authors also detected a molecular outflow in the core, thought to be due to a perturbed spiral structure in the inner disk.  \cite{1986ApJ...305..830B} also searched several galaxies for an H$_{2}$CO signal, discovering weak absorption in NGC\,660, while \cite{1998MNRAS.295..156P} looked at the H92$\alpha$ radio recombination line and found the dynamical centre is offset from the continuum peak.
More recently, H$_{2}$CO and NH$_{3}$ absorption have both been detected against the central region of the galaxy but with very different widths, suggesting that the two lines are probing different parts of the galaxy \citep{2013ApJ...766..108M,2013ApJ...779...33M}.

The H{\sc i} and CO distributions have been examined at low spatial resolution, as well as the optical and infrared properties, by \cite{1995AJ....109..942V} who found a flat disk rotation curve, a high average dust temperature similar to that of M82, but a low star-formation efficiency more similar to that found for non-starburst spirals.
The main gas-rich disk contains compact (on scales of 0\farcs2) radio continuum emission \citep{1990ApJ...362..434C}.
There is weak extended radio emission from the centre of the galaxy associated with the main disk, and observations with the VLA in C-array at 8.4\,GHz show many compact objects within the disk of NGC 660 \citep{2002ApJS..142..223F}.

Low angular resolution archive radio observations show NGC\,660 to be a fairly unremarkable starburst galaxy with no obvious active nucleus (AGN), and any radio emission from pre-existing compact sources appears to be purely starburst-related.
\cite{2005ApJ...620..113D} investigate a sample of IR-bright LINERs, including NGC\,660.  From the stellar velocity dispersion, they estimate the black hole mass in NGC\,660 to be log\,M$_{\rm BH} = 7.35$\,M$_\odot$, and calculate the Eddington ratio to be L$_{\rm bol}$/L$_{\rm Edd} = 4.0\times10^{-6}$, typical for LINERs and indicating a radiatively inefficient accretion flow.

Whilst several radio sources were seen in the 15-GHz map of \cite{1990ApJ...362..434C} (see also Figure \ref{images_archive}) no compact emission was seen at the dynamical centre of the galaxy.  At higher resolution, no emission from the galaxy was detected by \cite{1981ApJ...246...28J} in a three-station VLBI experiment at 18\,cm in 1978.
Archival MERLIN data from 1998 (Fig. \ref{images_archive}) show no evidence for a compact core at 5\,GHz, just a single compact ($\sim$50\,mas) source with an integrated flux density of 2.6 mJy, located 2.5 arcseconds away in the north-eastern end of the ring structure.

At radio wavelengths, the galaxy flared spectacularly sometime between 2008 and 2012 and has been reported to have developed a new continuum source with a GHz-peak spectrum and a peak flux density of $\sim$0.5\,Jy at 5\,GHz \citep{2013AAS...22115706M}.  High-resolution imaging has reportedly shown this increase to be due to a new source in the core, the nature of which is currently unclear.  Simultaneously, OH emission and absorption in both the ground state (18-cm) and excited (6-cm) lines has developed, and H$_{2}$CO absorption is seen against the new continuum source \citep{2013AAS...22115706M}.

Whatever the explanation for the outburst in this galaxy, the event is clearly unique.  In order to investigate its nature we examined archival X-ray and radio data, observed the galaxy with e-MERLIN at both 1.5 and 5\,GHz, and at 1.4\,GHz with the EVN, obtaining both a high angular resolution continuum image to look for evidence of small-scale structures, and spectral line data to probe the H{\sc i} and look for evidence of outflowing gas; this paper describes these observations.  Section \ref{section_obs} outlines the observational setups used to obtain new data and the archival data which has been examined, sections \ref{section_continuum} and \ref{section_spectra} describe the continuum and spectral line results, we explore possible scenarios in \ref{section_discussion}, and summarise our findings in section \ref{section_summary}.


\section{Observations}\label{section_obs}

NGC\,660 was observed at 1.4\,GHz with 11 stations of the EVN (Effelsberg, Westerbork, Lovell, Onsala, Medicina, Noto, Torun, Svetloe, Zelenchuk, Badary, Urumqi) on 30 October 2013 (project code EA054).  The observation used 3C84 and 3C120 as bandpass calibrators, and the nearby source J0143+1215 as a phase calibrator with a switching cycle of seven minutes on NGC\,660 to three minutes on the phase calibrator.
The phase calibrator has a position of RA\,$=01^{\rm h}43{\rm m}31\rasec092221$ Dec\,$=+12\degr15'42\farcs93343$ with positional uncertainties of 1.02 mas in RA and 0.78 mas in Dec.
The data were correlated in two passes, one pass of 8 sub-bands (each with 32 channels, dual polarisation and 2-second integrations) for continuum, and one pass with one high-frequency resolution sub-band (1024 channels, dual polarisation, 2-second integrations) to provide spectral line information with a velocity resolution of $\sim$3.3\,km/s.
No serious problems were reported during observations, however Noto recorded no fringes in the line pass, and several antennas suffered from significant interference.
Data reduction proceeded according to the EVN User Guide, including: interference rejection, fringe fitting, ionospheric corrections, phase calibration, amplitude calibration and bandpass calibration.  Both the line and continuum datasets were processed in the same way, up to the bandpass stage.  Once calibrated, the continuum data were averaged and imaged, whilst each channel of the line data was imaged individually to form a cube for spectral analysis.
Self-calibration was carried out on both the phase calibrator and target, initially in phase-only mode, then using both amplitude and phase.
Maps were made during each iteration of self-calibration, and the solutions were carefully inspected at each step.

Lower angular resolution e-MERLIN observations were also made in May/June 2013 at both 1.4 and 5\,GHz.  These observations used all six telescopes of the standard array (Mk2, Pickmere, Darnhall, Knockin, Defford, Cambridge) and used the same phase calibrator as the EVN observations.  For flux and bandpass calibration, 3C286 and OQ208 were observed using the same configuration.  The data were processed using v0.7 of the e-MERLIN pipeline \citep{pipeline}, with further editing and self-calibration cycles carried out by hand.


\subsection{Archival observations}

MERLIN 1.4- and 5-GHz observations from 1998 have been extracted from the archive and re-processed, and two Very Long Baseline Array (VLBA) observations taken in 2001 (BF068; 5.0\,GHz) and 2010 (BC191; 8.4\,GHz) have also been examined.  The 5-GHz 2001 VLBA data cover a total of $1^{\rm h}23^{\rm m}$ split into two observations, separated by four hours, using four 8-MHz sub-bands each split into 16 channels.  The 8.4-GHz 2010 data consist of two observations of eight minutes each, separated by 96 days, each observation was made using 8 sub-bands of 16\,MHz split into 128 channels each.
In addition, the VLA A-array 1986 15\,GHz data from \cite{1990ApJ...362..434C} has also been extracted and analysed.

Four separate X-ray observations from the {\it Chandra} archive were used, taken in January 2001, February 2003 and, December and November 2012
(ObsIDs 1633, 4010, 15333 and 15587). The data consist of a total of $\sim$52\,ks of exposure time, $\sim$7\,ks before the radio outburst and $\sim$45\,ks after. All of these data were obtained using the Advanced CCD Imaging Spectrometer (ACIS-S). The data were processed using {\sc CIAO} v4.5, with the most recent calibration files \citep{2006SPIE.6270E..60F}. Spectra covering 0.1-10\,keV were extracted using a circle of radius 6 arcsec centred on the location of the radio nucleus, but avoiding the point source to the west of the nucleus (see Section \ref{sec_xray}). Background spectra were extracted using an annulus surrounding the source extraction region, with an outer radius of 25 arcsec and an inner radius of 13 arcsec, avoiding nearby X-ray point sources.


\section{Continuum components}\label{section_continuum}

\begin{figure*}
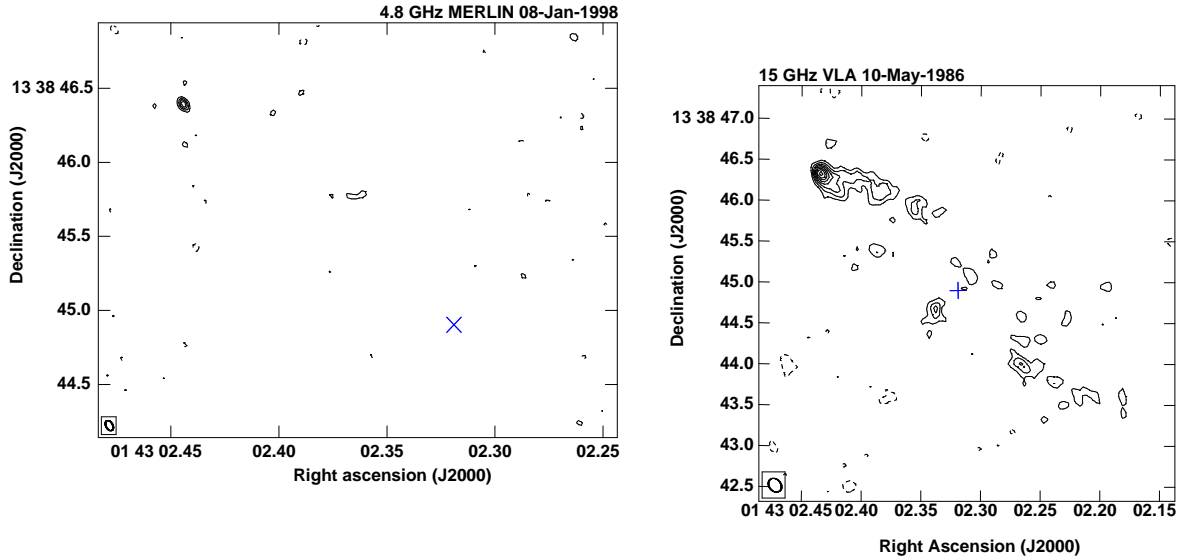

\centering
\begin{tabular}{cc}
\includegraphics[width=7cm,angle=270,origin=c]{merlin+stars.ps} &
\includegraphics[width=7cm,angle=0,origin=c]{N660_VLA_8.4GHz_J2000.ps} \\
\end{tabular}
\caption{\label{images_archive} MERLIN archive data at 4.8\,GHz from an observation on Jan 9$^{\rm th}$ 1998 (left), and VLA A-array 15-GHz data from 1986 (right; \protect\citealt{1990ApJ...362..434C}).  The only radio source visible in the MERLIN data is a supernova remnant, located towards the north-east end of the main disk of the galaxy.  In the VLA data from \protect\cite{1990ApJ...362..434C} the supernova remnant is also present, along with some emission from the central star-forming ring.  In both images the cross marks the position of the new continuum source reported here.  In both maps the peak flux is 1.7\,mJy/beam; the contours are set at (-1, 1, 2, 3, 4, 5, 6, 7, 8, 9, 10) $\times$ 345\,$\mu$Jy/beam ($3\sigma$) in the MERLIN map, and (-2, 2, 3, 4, 5, 6, 7, 8, 9, 10) $\times$ 180\,$\mu$Jy/beam in the VLA map.}
\end{figure*}

\subsection{Radio results: a time-line}

No central compact radio source was detected within NGC\,660 in either archival MERLIN observations from 1998 (to a 3-$\sigma$ limit of 0.3\,mJy/bm, Figure \ref{images_archive} left), or VLBA observations taken in 2001.  The archival VLBA observations in September 2010 show that a single compact source (beam size 2$\times$1\,mas) was present at the centre of the galaxy at 8.4\,GHz by late 2010.  The peak flux density rises from 18.3$\pm$0.9\,mJy/beam on September 16th to 36.0$\pm$1.8\,mJy/beam on December 21st 2010.

The two e-MERLIN continuum datasets from mid-2013 both show a strong compact source at the same position as the 2010 VLBA detection, unresolved in both frequency bands (150\,mas at 1.4\,GHz, 50\,mas at 5\,GHz).
In the 2013 e-MERLIN data, the integrated flux density in the centre of the 5-GHz band is 364$\pm$18\,mJy.  Comparing this with the MERLIN pre-outburst limit results in a factor of $>$1200 change in brightness.  At the centre of the e-MERLIN band observed at 1.4-GHz, the integrated flux density is 198$\pm$9.9\,mJy, giving a fairly flat overall spectral index of $\alpha = 0.5$, where $S \propto \nu^{\alpha}$. 

\begin{figure}
\includegraphics[width=8cm,angle=0]{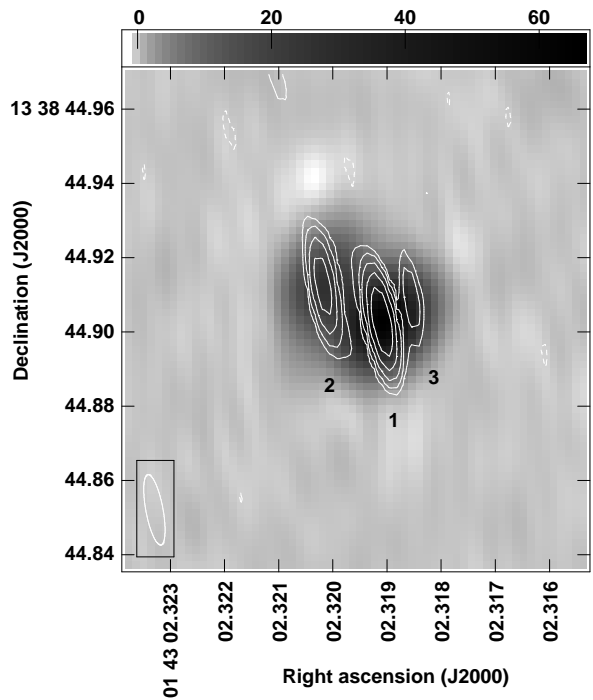}
\vspace{0.3cm}
\caption{\label{fig_continuum} Continuum image of the new source in NGC\,660 showing the jet-like components.  The ``low" resolution (natural weighting; 26$\times$20\,mas) image is shown in grey scale, with contours from the ``high" resolution (uniform weighting; 19$\times$6\,mas) image overlaid.  The numerals correspond to those of the spatial fit parameters listed in Table \protect\ref{table_cont}.  The data are contoured at (-1, 1, 2, 4, 8, 16, 32, 64) $\times$ 1.736\,mJy/beam, the greyscale range is -1.26 to 66.49\,mJy/beam.}
\end{figure}

The most recent observation, the October 2013 EVN data (with an angular resolution of 26$\times$20\,mas), shows an obvious jet-like feature to the north-east of the central brightest source (Figure \ref{fig_continuum}, greyscale).  When the EVN data are imaged with uniform weighting, giving greater resolution (19$\times$6\,mas) at the expense of increased noise, there also appears to be a weaker feature to the west (Figure \ref{fig_continuum}, contours).  The high resolution image shows a compact central object with an integrated flux density of 69.1$\pm$3.5\,mJy.
Despite having greater angular resolution, the 2010 VLBA data shows no sign of the structures seen in the EVN data taken three years later, only the central brightest source.  This could be because the object is evolving spatially in time, or simply due to the lack of sensitivity and poor $u-v$ coverage of the earlier VLBA snapshot observations.

Between e-MERLIN and EVN scales there is a significant decrease in the flux detected.  Summing the fitted components in the EVN model gives a total flux density of $\sim$105$\pm$5\,mJy, while the e-MERLIN flux density measured in the sub-band closest to the EVN measurement is 196$\pm$10\,mJy, almost twice that detected by the EVN.  This large difference in flux is likely due to a combination of the two arrays not being sensitive to the same spatial scales, and the source possibly fading as it evolves in the five months between the observations.

Table \ref{table_cont} shows the results of two-dimensional Gaussian fits to the EVN continuum images.  The two-component fit was carried out on the naturally-weighted image, while the three-component fit is for the higher resolution uniform-weighted image.
If component 2 is, as it appears to be, moving, then we can estimate limits on its velocity.  Assuming that the event occurred between 2008.0 \citep{2013AAS...22115706M} and 2010.7 (epoch of the first of the VLBA detections), then the apparent speed of component 2 is $0.6c \leq v \leq 1.2c$, relative to component 1.


\subsection{Archival X-ray results}\label{sec_xray}

\begin{figure}
\centering
\includegraphics[width = 8cm,trim = 0 0 0 0, clip=true]{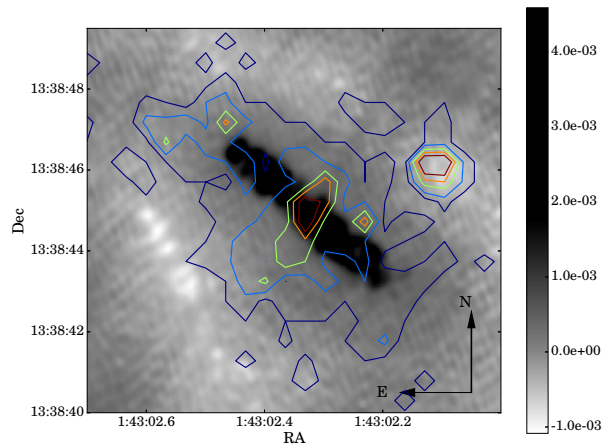}\hspace{1pt}
\caption{Combined radio (archive 1998 MERLIN 1.4-GHz; greyscale) and X-ray ({\it Chandra} 0.1-10\,keV; contours) image of the diffuse emission around the nucleus of NGC 660. The X-ray contours use the combined data from the two observations from 2012, after the radio outburst. The contour levels are at (4.5, 12.5, 22.5, 30.0, 37.5) $\times 10^{-5}$ counts $\mathrm{s^{-1}}$.  The greyscale range is from -1.0 to +4.6\,mJy/beam.  The X-ray emission is well aligned with the radio emission on this scale, which is extended in the direction of the main disk of the galaxy.}
\label{radio+xraycontour}
\end{figure}

The X-ray image of the nucleus after the radio outburst (0.1-10\,keV, Fig. \ref{radio+xraycontour}) shows a region of diffuse emission around the position of the compact radio core. The emission is extended $\sim$8 arcsec to the north-east of the core and $\sim$4 arcsec to the south-west, coincident with both the orientation of the main disk of the galaxy, and the jet-like components seen in the high-resolution radio data.  The extension seen perpendicular to the main disk is not aligned with the axis of the star-formation ring and may be associated with a starburst wind like that seen in M82 \citep{1997A&A...320..378S}.  There is also a peak in the emission at the position of the radio core, which is 2-3 times brighter than the surrounding emission, but it is unclear whether this is a brighter patch of diffuse emission or is actually a peak of emission from the nucleus.

Unfortunately, the data taken before the outburst are not sufficient to make a meaningful image, as the very low count rate ($\mathrm{1.30\pm 0.10\times 10^{-2}} {\rm s}^{-1}$) meant that the exposure time was insufficient.
The 0.5-10 keV X-ray fluxes (dependent on the model used) were 1.24$^{+0.37}_{-0.54}$$\times$10$^{-13}$ ergs\,cm$^{-2}$\,s$^{-1}$ and 1.85$^{+0.19}_{-0.16}$$\times$10$^{-13}$ ergs\,cm$^{-2}$\,s$^{-1}$ before and after the burst, respectively.  Assuming a redshift of 0.002842, this corresponds to luminosities of 2.20$^{+0.72}_{-0.50}$$\times$10$^{+39}$ ergs\,s$^{-1}$ and 3.28$^{+0.32}_{-0.28}$$\times$10$^{+39}$\,ergs\,s$^{-1}$ respectively, so potentially brighter afterwards.

\begin{table}
\begin{center}
\begin{tabular}{cccccc}
		& RA 			& Dec			& Peak		& Int.		& T$_{\rm B}$		\\
		& (J2000)		& (J2000)		& (mJy/bm)	& (mJy)		& ($\times 10^{7}$K)	\\
\hline
1		& 02.319		& 44.9033		& $75.4\pm3.8$	& $80.0\pm4.0$	& $>$7.2	\\
2		& 02.320		& 44.9130		& $26.3\pm1.3$	& $25.6\pm1.3$	& $>$2.3	\\
\hline
1		& 02.319		& 44.9028		& $62.8\pm3.1$	& $69.1\pm3.5$	& $>$11.6	\\
2		& 02.320		& 44.9127		& $24.9\pm1.2$	& $27.7\pm1.4$	& $>$4.6	\\
3		& 02.319		& 44.9068		& $10.1\pm0.5$	& $10.6\pm0.5$	& $>$1.8	\\
\hline
\end{tabular}
\caption{\label{table_cont} Results of 2-component (top) and 3-component (bottom) fits to the naturally-weighted  and uniform  continuum images respectively.  Labels correspond to those in Figure \protect\ref{fig_continuum}.  Positions are relative to $01^{\rm h}43^{\rm m}$ $13\degr38'$, and phase referenced to J0143+1215 (RA $= 01^{\rm h}43^{\rm m}31\rasec092221$ Dec $= 12\degr15'42\farcs93343$; J2000) which has a catalogued positional uncertainty of $\pm1.02$\,mas in RA and 0.78 mas in Dec.  Brightness temperatures are calculated using the beam major axis and are lower limits only.}
\end{center}
\end{table}

\section{Spectral features}\label{section_spectra}

\subsection{Radio H{\sc i} spectra}

\begin{figure}
\includegraphics[width=6cm,origin=c,angle=270]{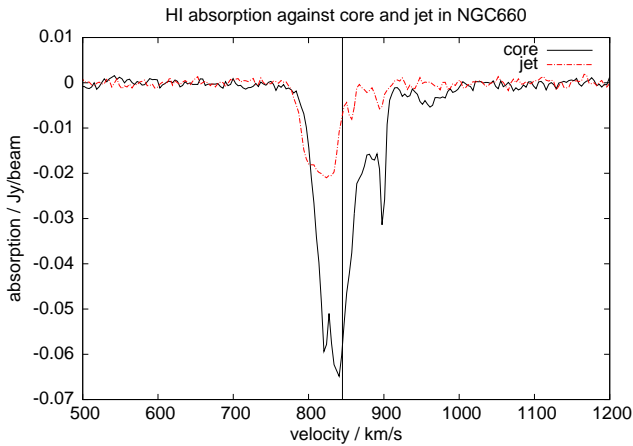}
\caption{\label{fig_spectrum} H{\sc i} spectrum against components 1 (black, solid) and 2 (red, dashed) from the October 2013 EVN observation.  The vertical line marks 845\,km\,s$^{-1}$ the systemic velocity of the galaxy.}
\end{figure}

Figure \ref{fig_spectrum} shows two radio spectra taken from the EVN line dataset.
The line pass has a higher noise level than that of the continuum dataset, with the result that it was not possible to make a reliable spectrum of source 3, the weakest of the three sources shown in Figure \ref{fig_continuum}.  In Figure \ref{fig_spectrum}, the solid black line is from source 1, while the dashed red line is from source 2 (see Figure \ref{fig_continuum}).  It is clear that both radio spectra are composed of several narrow spectral components, with significant differences between the two.

\begin{figure}
\includegraphics[width=8cm]{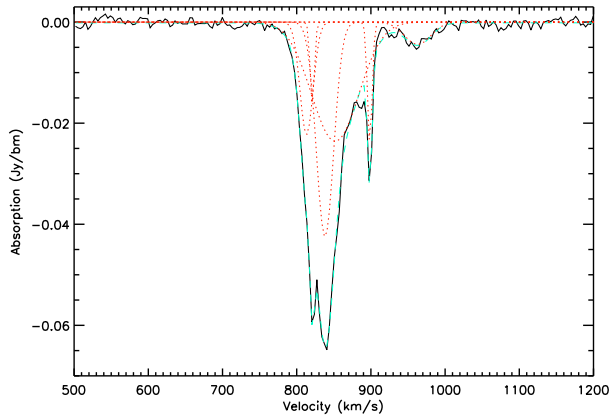}
\caption{\label{fig_fit} Spectrum against component 1 showing the six Gaussian components from the fit in Table \ref{table_spectra} (red dotted lines).  The composite fitted spectrum is also overlaid (pale dashed line).}
\end{figure}

The strongest component in the spectrum of the core is very close to the systemic velocity (845$\pm$1\,km\,s$^{-1}$; \citealt{1993ApJS...88..383L}).  The spectrum from source 2 is also composed of several narrow velocity components, but with the features on the blue side of the systemic velocity significantly shifted to lower velocity, the same sharp peak near 900\,km\,s$^{-1}$, and no broad feature near 960\,km\,s$^{-1}$.
Attempts were made to fit the complex line structure with a range of multi-Gaussian component models. The best constrained of these models comprised of six Gaussian components, the results of which are presented in Table \ref{table_spectra} and shown in Figure \ref{fig_fit}.  The reduced $\chi^2$ for this fit is poor (9$\times$10$^5$), but it should be noted that excluding any of these components (or adding additional Gaussians) made the quality of the fit significantly worse.
If the new compact radio sources are, as they appear to be, at the centre of NGC\,660, then each absorption profile is probing a very narrow line of sight through the near-side of the galactic disk, with each of the narrow spectral components corresponding to a different physical feature along the line of sight.

The EVN spectrum taken against the brightest source (component 1) matches very well with that seen in recent Westerbork observations (van Bemmel et al in prep).  However there is less absorption in the red-shifted wing centred around 960\,km\,s$^{-1}$ than is seen in the spatially-unresolved Westerbork spectrum, and no evidence of the blue-shifted wing below 850\,km\,s$^{-1}$ seen in lower spatial resolution data.  
This is not surprising, since the EVN spectrum is probing a much narrower line of sight through the galaxy than that of the much larger Westerbork beam.
Significantly, the deep absorption seen on the blue side of the systemic velocity in the spectrum taken against source 2 is systematically $>5$\,km\,s$^{-1}$ ($>$4.8$\sigma$) bluer than that taken against the central radio source.  It is likely that this is not purely due to the large-scale disk rotation, as \cite{1992ApJ...401..508B} found a gradient of 1.58\,km\,s$^{-1}$\,pc$^{-1}$ across the disk in H{\sc i} absorption; this would result in a difference of just 1.9\,km\,s$^{-1}$ between the positions of sources 1 and 2.

The EVN H{\sc i} spectrum shows that the peak 21-cm opacity, $\tau$, is 1.58, $\int \tau dv = 66$ \,km\,s$^{-1}$ and hence N$_{\rm HI} = 0.182\times10^{21} (T_{S} / 100K) \int \tau dv = 12.1 \times 10^{21}$ cm$^{-2}$ where we assume T$_{S}$ = 100K.  Compared to the results of \cite{1999ApJ...524..684G} on Seyfert and starburst galaxies, this is quite high.
In NGC\,660 \cite{1992ApJ...401..508B} find the H{\sc i} column density N$_{\rm HI}$/T$_{S}$ to be $7.4\times10^{19}$ cm$^{-2}$/K against their component 4, the closest to the core, although this is measured with the much larger 1\farcs4 beam of the Very Large Array in A-configuration.

\subsection{X-ray spectral fitting}

\begin{figure}
\centering
\includegraphics[width = 8cm,trim = 0 0 0 0, clip=true]{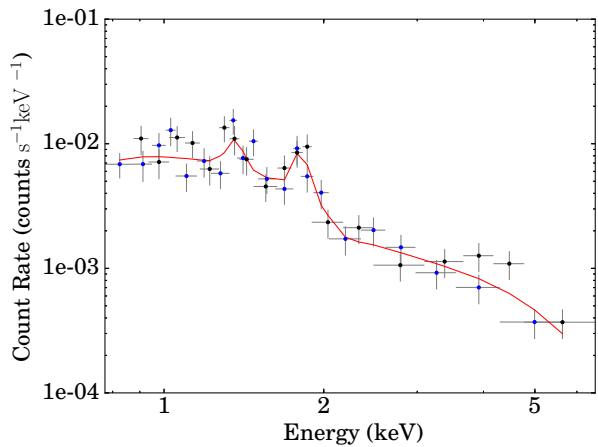}\hspace{1pt}
\caption{X-ray spectra from two separate {\it Chandra} observations taken after the radio outburst (red and blue points). The line shows a model fitted to the data, consisting of a power law and two Gaussians. Both spectra show strong emission lines at approximately 1.80 and 1.35 keV, which are most likely due to silicon and magnesium transitions.}
\label{xrayspectrum}
\end{figure}

The low X-ray count rate meant that it was not possible to extract a useful spectrum from the core region alone. The X-ray spectrum of the entire region of diffuse emission around the nucleus was extracted from the data taken after outburst (Fig. \ref{xrayspectrum}). It possesses a continuum which can be fairly well fitted by a power law with a spectral index of $1.61^{+0.13}_{-0.12}$ (reduced $\chi^2 = 1.26$).

However, there are several large deviations from this, which coincide with well-known emission lines.
These can be well fitted by Gaussians at $1.80^{+0.04}_{-0.04}$\,keV and $1.35^{+0.04}_{-0.03}$\,keV, giving a total reduced $\chi^2$ of 1.0 together with a power law with spectral index $1.59^{+0.14}_{-0.14}$. The widths of these lines, $5^{+10}_{-5}$\,eV and $0.1^{+80}_{-0.1}$\,eV respectively, are not well constrained due to the low number of counts in the spectra and consequent low spectral resolution. The Gaussians are, however, required to account for the excesses in the data and provide a good fit. The energies of the Gaussians are consistent with emission lines of magnesium and silicon.
The features are both seen in both post-outburst spectra, and are therefore unlikely to be artefacts.
These emission lines are commonly seen in supernova remnants and are therefore likely to be associated with the star formation observed in this region \citep{2007MNRAS.374.1290G}.  Indeed, emission lines from these elements have also previously been seen in other wavebands \citep{2004A&A...414..825S}.

The {\it Chandra} spectra can also be reasonably well-fitted with a model for shocked gas which can be used describe supernova emission ({\it vpshock}, reduced $\chi^2 = 1.23$), however the parameters of the best-fitting models require either a high gas temperature ($\sim$30\,keV) or very high abundances of some elements (tens of times higher than solar abundance), due to the strength of the lines in the spectra. As would be expected, the best fitting parameters require high abundances of silicon and sulphur at low temperatures, as well as helium and calcium. As with the images, the pre-outburst spectra are of too poor quality to perform meaningful analysis on, but are not inconsistent with possessing strong emission lines.

\begin{table}
\begin{center}
\begin{tabular}{ccccc}
Component	& Depth		& Width		& Velocity	& $\tau_{\rm peak}$	\\
		& Jy		& km\,s$^{-1}$	& km\,s$^{-1}$	& 	\\
\hline
1		& -0.022	& 8.7		& 813		& 0.32	\\ 
2		& -0.016	& 2.9		& 821		& 0.22	\\ 
3		& -0.043	& 11.7		& 838		& 0.77	\\ 
4		& -0.024	& 32.2		& 851		& 0.36	\\ 
5		& -0.024	& 3.2		& 898		& 0.36	\\ 
6		& -0.0045	& 18.3		& 963		& 0.06	\\ 
\hline
1		& -0.017	& 10.2		& 800		& 0.24	\\ 
2		& -0.008	& 6.3		& 816		& 0.11	\\ 
3		& -0.020	& 9.7		& 830		& 0.29	\\ 
4		& -0.008	& 4.0		& 857		& 0.11	\\ 
5		& -0.005	& 6.8		& 894		& 0.06	\\ 
6		& -0.0007	& 18.3		& 984		& 0.009	\\ 
\hline
\end{tabular}
\caption{\label{table_spectra} Resulting parameters of 6-component fits to the radio spectra (Figure \ref{fig_spectrum}) for component 1 (top) and component 2 (bottom).  The systematic uncertainty on the velocity measurements is $\pm1.6$\,km/s.}
\end{center}
\end{table}


\section{The nature of the event}\label{section_discussion}

While many radio-loud quasars and galaxies do show some degree of time-variable behaviour across the electromagnetic spectrum, the ongoing event in NGC\,660 is extreme.  The spectacular and sudden increase in radio emission, together with the appearance of new radio structure, has been suggested to originate in either the onset of nuclear activity, a tidal disruption event involving the central black hole (with precession possibly explaining some of the morphology), or a highly luminous supernova event.  At Galactic coordinates (141,-47), the likelihood of a foreground Galactic flaring event is low, and something this bright, so far from the Galactic plane, would be unlikely to have escaped attention.  The high-resolution radio structure of this source presented here apparently shows a core with two lobes, similar to the core-jet morphology often seen in AGN.  One likely explanation is that the galaxy is undergoing a burst of enhanced nuclear activity related to the recent merger or close encounter event which created the polar ring and caused the starburst activity.
In this section we compare the information obtained from these new observations and archival data with those of possible explanations, some more likely than others.


\subsection{A central starburst}

The central compact EVN component has a brightness temperature of $T_b>10^7$\,K (see Table \ref{table_cont}).
Given that typical central starbursts have $T_b\leq10^5$\,K, it is unlikely that an active central starburst is a viable explanation for the source.  Also, since the new source has appeared since the VLBA observations in 2001, and likely since the last reported non-detection of flaring activity in 2008, the measured brightness of the source would imply some 10 to 100 ``normal" radio supernovae have occurred in a very small region ($\sim$1.1\,pc) over a handful of years.  This seems unlikely, especially when compared to known highly active starbursts.  One of the most powerful nearby starbursts is that in Arp\,220.  Here \cite{2006ApJ...647..185L} find the radio supernova rate to be 4$\pm$2 per year across two nuclei each some 100 to 150\,pc across, separated by 370\,pc. 
At 13\,Mpc, 1" is 63\,pc, so we would easily expect to resolve individual SN in NGC\,660 if this were the explanation, unless several tens of RSNe occurred within a region of $\sim$1\,pc. 
The spectral features detected in the {\it Chandra} data do show the presence of chemicals often seen in supernova remnants, although their likely presence in the pre-outburst spectra strongly suggest that these lines are associated with the pre-outburst star formation observed in this region, rather than with new activity.


\subsection{A single luminous radio supernova/remnant}

Another possibility is that the outburst is due to a single stellar-scale explosive event, although the distance of NGC\,660 would make this significantly more luminous than most known radio supernovae or supernova remnants.  The 1.4-GHz power from the EVN observation is $1.42 \times 10^{21}$\,W\,Hz$^{-1}$; the fact that this is three orders of magnitude more powerful than known evolved supernova remnants in our own Galaxy such as Cas A ($10^{18}$\,W\,Hz$^{-1}$; \citealt{1977A&A....61...99B}), along with the fact that the object was no more than six years old at the time of observation, makes this explanation very unlikely.  The highly luminous radio supernovae in Arp 220 are a more promising comparison, although even here the source in NGC\,660 is an order of magnitude more powerful \citep{2006ApJ...647..185L}.

The extremely energetic GRB-associated supernovae SN1998bw and SN2009bb are closer in luminosity to this event and resulted in relativistic jets.  Both were type Ib/c events and relatively short-lived in the radio \citep{1998Natur.395..663K, 2010ApJ...725....4B}; the time-scale of the event in NGC\,660 is more characteristic of a longer-lived type II event.  One of the brightest RSN ever, SN1986J, has developed a complex and evolving morphology, and a spectrum which flattens over time \citep{2010ApJ...712.1057B}, however the lower limit on the expansion speed for the source in NGC\,660 is significantly greater than the expansion seen in 1986J, or other nearby ordinary RSN which have been well-studied on VLBI scales (e.g. \citealt{2002ApJ...581..404B, 2010MNRAS.408..607F, 2011A&A...535L..10M}).

A further comparison can be made with the unusual radio sources in M82.  One suggestion for the origin of the fading source 41.95+575 is a historical GRB seen off-axis from our perspective.  This source is resolved on VLBA scales and would have had a peak luminosity some two orders of magnitude greater than the event in NGC\,660, but the VLBI-detected components are expanding at only $\sim$1800\,km/s \citep{2005MmSAI..76..586M}.


\subsection{A tidal disruption event}

A handful of optically non-active galaxies have displayed X-ray flaring behaviour interpreted as disruption events \citep{2002RvMA...15...27K}.  The characteristic behaviour of such events includes huge peak X-ray luminosity, variability amplitudes of up to a factor of $\sim$200 with a short rise time and declining over months to years, and an ultra-soft X-ray spectrum.  In the case of NGC\,660, the expected X-ray peak luminosity should have exceeded the flux of the diffuse continuum by several orders of magnitude, but the archival X-ray data does not show this expected huge peak in luminosity.

An example of an X-ray flare thought to be due to a stellar disruption event is the case of NGC5905.  An X-ray flare was detected in 1991 and VLA observations were carried out to search for post-outburst radio emission in 1996, but no compact central object was detected putting a limit of $<10^{20}$\,W\,Hz$^{-1}$ on the radio luminosity \citep{2001astro.ph..6422K}. 
On e-MERLIN scales we see a source appear in NGC\,660 compared to archival data at the same resolution, implying a variability factor of $\sim$1200 in the radio.  However the rise time is unknown, and the decay is, as yet, not well-sampled.  There is no evidence for prior low-luminosity AGN activity, with no optically-hidden radio AGN present in archival radio data.

The case of Swift J1644+57 also provides an interesting comparison.  Displaying a mildly relativistic outflow and spectral features due to synchrotron and inverse Compton processes, \cite{2011Sci...333..203B} draw an analogy with a temporary small-scale blazar-like event.  \cite{2011MNRAS.416.2102G} discuss transient radio emission from such events and model the resulting characteristics, focussing on off-axis events.  In the case of NGC\,660, we have no information on the time-scale from outburst to peak luminosity, and information on the rate of decay will only come with time.


\subsection{A new period of AGN activity}\label{section_llagn}

Whilst low-power jets from tidal disruption events are short-lived and dissipate their energy in to the ISM rapidly, jets from AGN last orders of magnitude longer and propagate to much larger distances.  The most likely scenario is that the central super-massive black hole is beginning a period of activity, turning NGC\,660 from a LINER-type galaxy into a low-luminosity AGN.

There is now much evidence that a significant fraction of nearby LLAGN galaxies host flat-spectrum compact cores with associated parsec-scale jets (e.g. in M51 Rampadarath et al in prep; more generally in \citealt{2002A&A...392...53N, 2005A&A...435..521N, 2013MNRAS.432.1138P, 2014ApJ...787...62M}).  In NGC\,660, the central source detected by e-MERLIN is both compact and flat spectrum, and the EVN map shows evidence of a core-jet structure.  The spectral index of the EVN sources cannot be calculated in this case, since there is only 1.4-GHz information available at present.
The EVN morphology resembles that of a compact symmetric object.  These are the youngest class of AGN, with steep spectra, measurable expansion and the lobes entirely contained within the (clumpy, asymmetric) medium of the host galaxy.  Such activity is generally triggered by large-scale collision or merger events driving material into the central region of the galaxy, and NGC\,660 has undergone a relatively recent encounter which greatly disturbed its morphology.

\begin{figure}
\centering
\includegraphics[width=6cm,angle=270]{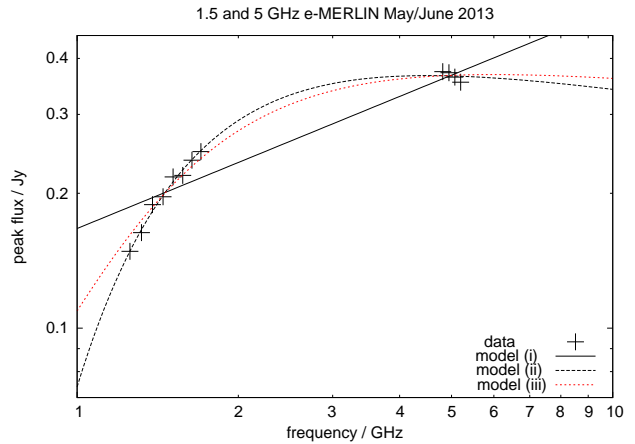}
\caption{\label{fig_fluxes} Peak flux measurements per sub-band from e-MERLIN observations made at 1.5 and 5\,GHz in May/June 2013 ($+$).  The lines shows least squares fits to the e-MERLIN data, with solid black, dashed black and dotted grey lines corresponding to models (i), (ii) and (iii) in the text (see Section \protect\ref{section_llagn}).}
\end{figure}

Whilst limits placed on the apparent expansion velocity by the archival data suggest that the motion could be mildly relativistic, if components 2 and 3 are a jet and counter-jet respectively, then the ratio of their flux densities ($\sim$2.5) suggests that they are not moving relativistically. 
Further EVN observations have been requested in order to look for morphological changes or relative motion of the components.  If this source is a newly-formed jet, these observations will enable us to better constrain the expansion velocity, resulting in a better estimate for the onset of activity.

As discussed in section \ref{section_continuum}, the average spectral index of the new radio component is $\alpha = 0.5$.  With the wide bandwidth of the new e-MERLIN system it is possible to measure the spectral index within each band; Figure \ref{fig_fluxes} shows the flux measurements for the new (unresolved on e-MERLIN scales) component in each of the eight sub-bands at 1.4\,GHz and four sub-bands at 5\,GHz.  The results clearly show that this new radio object is a GHz-peaked spectrum (GPS) source, with a synchrotron spectrum above a turnover at around 3$-$3.5\,GHz.
The spectral index below the peak is $\alpha = +1.6$, while above the peak it is $\alpha = -0.72$.  From a large sample of GPS sources, \cite{1997A&A...321..105D} found that (using the same convention $S\propto\nu^{\alpha}$) the average spectral index on the optically thick side of the turnover frequency is +0.51, whilst that on the optically thin side is $-$0.73.  NGC\,660 is steeper on the optically thick side than this average, suggesting stronger self-absorption, although there is a large scatter in the sample of \cite{1997A&A...321..105D}.

GPS sources generally represent a very early phase in the evolution of radio galaxies, with their extent entirely contained within the central kpc of their host galaxy, morphologies similar to those of larger more evolved classical double-lobed radio galaxies, and the peak in their spectra likely due to synchrotron self-absorption (\citealt{1998PASP..110..493O} and refs therein).  This description closely matches the observed properties of the $\sim$6-year old source in NGC\,660.
A young source would be expected to have a steep, constant spectral index, while older AGN-related structures (jets, lobes) flatten at low frequency due to radiative losses.  In the case of NGC\,660, the EVN observations clearly show multiple source components within the e-MERLIN beam.

Fitting the e-MERLIN spectral points with (i) a simple power law, (ii) a power law with free-free absorption by a foreground screen of ionised gas, and (iii) a self-absorbed Bremsstrahlung spectrum, we find that none of these models are a good fit to the data, all being less steep at frequencies above the turnover (Fig. \ref{fig_fluxes}).  Further e-MERLIN measurements have been requested, covering a wider frequency range at C-band (4$-$8\,GHz), to better constrain the fits, but the explanation could simply be that the different compact components seen with the EVN have different spectra, an effect also discussed by \cite{1997A&A...321..105D} in the context of compact double sources.

The X-ray spectra of the emission region after the outburst show a total flux of $1.85^{+0.19}_{-0.16}\times 10^{-13}$\,ergs\,cm$^{-2}$\,s$^{-1}$.  Calculations using the method from \citet{1983ApJ...264..296M} show that, if the radio emission seen at the nucleus is due to synchrotron self-absorption, the X-ray flux expected from the region is negligibly small for an assumed radio nucleus size of  $\sim$0.5\,mas, from the VLBA observations. All of the parameters in this calculation are either well constrained or do not have a significant impact on the final value, except for the source size.  Since the core itself is unresolved, the physical size could be smaller.  A typical core size of 0.2 mas would still produce only very low X-ray fluxes. A size of $\sim$0.1\,mas or smaller, which is the case in only a small fraction of AGN \citep{2004ApJ...616..110H}, is needed to produce a flux comparable to the $10^{-13}$\,erg\,cm$^{-2}$\,s$^{-1}$ diffuse flux. If the source is a newly-awoken AGN, X-ray emission from additional processes to synchrotron self-Compton are likely to be present, however, many examples of low luminosity AGN with X-ray luminosities of $<10^{41}$\,erg\,s$^{-1}$ exist, equating to a flux of $<10^{-15}$\,erg\,cm$^{-2}$\,s$^{-1}$ at the distance of NGC 660 \citep{2003ApJ...583..145T}. Furthermore, the radio spectrum indicates a very high absorbing column in the nucleus, which would cause a large degree of X-ray absorption if covering the X-ray source. It is therefore not unexpected that no strong X-ray nuclear emission is observed, even if a new AGN has appeared. 
If the increase in flux is due to a new period of SMBH activity in the centre of NGC\,660, then the radio luminosity measured here places it relatively low on the fundamental plane \citep{2003MNRAS.345.1057M}, hence the X-ray luminosity would also be expected to be low and the non-detection is consistent.


\section{Conclusions}\label{section_summary}

The outburst in NGC\,660 has been explored using archival radio and X-ray data, together with new observations at 1.4 and 5\,GHz with e-MERLIN, and at high resolution with the EVN at 1.4\,GHz.  Our observations show the appearance of a bright new source in the core of the galaxy, an object not present in any archival radio images of the galaxy prior to 2010.  This new object is compact on a scale of $\sim$50 milliarcseconds with e-MERLIN, but displays an obvious jet-like component to the north-east and a weaker possible counter-jet to the west at higher angular resolution with the EVN.  The evidence from these new radio observations, combined with archival data from {\it Chandra}, MERLIN, the VLA and the VLBA, suggests that we are seeing the activation of a low-luminosity AGN in NGC\,660.

H{\sc i} spectra obtained during the EVN observation show strong absorption along this new line of sight through the disturbed disk of the galaxy, potentially absorbing any associated X-ray emission.  The H{\sc i} spectrum of the new central radio source strongly resembles that seen at lower resolution with Westerbork-only data, post-outburst, with several narrow velocity components due to absorption in the disk, and a slightly broader red-shifted wing suggesting an inflow of material along our line of sight.

The EVN flux point at 1.4\,GHz is almost 50\% lower than that measured by e-MERLIN at the same frequency 130 days earlier.  This is likely due to a combination of overall fading of the source, spectral evolution, and resolution effects between the two arrays; the relative effects of these two explanations will be tested by planned follow-up observations.
The astrometric capability of the EVN will also allow us to better constrain the direction of expansion and the velocity of the components, resulting in a better estimate for t$_0$; an apparent velocity of 1.2\,c would result in a movement of 0.4\,pc over one Earth-year, a shift of 5.8\,mas at a distance of 13\,Mpc.
There is also much information to be extracted from further e-MERLIN observations at lower angular resolution.  Observations at both 1.4 and 5\,GHz have been requested in order to measure any changes to the radio spectral energy distribution of the new source, and additionally probe several additional spectral lines, providing valuable information on physical conditions of the gas in the centre of NGC\,660.

Whatever the cause of the increase in radio luminosity, the fact that such variability does occur in LINER-type galaxies is potentially a serious issue for statistical studies of the link between star formation, fuelling and accretion in low luminosity AGN.
If this is, as it appears to be, a new period of AGN activity, then in the short term we can expect to track the expansion as the jets move out through the galactic medium, and watch for evolution of the GPS-nature of the spectrum.
In the much longer term, if strong enough the outflows could quench future star formation by clearing out the gas reservoirs needed.
GPS sources are seen across a wide variety of peak luminosities, galaxy types and redshifts, and are useful probes of the early universe.  NGC\,660 is now one of the closest examples, and presents an ideal opportunity to watch as it evolves.


\subsection*{Acknowledgements}

The European VLBI Network is a joint facility of European, Chinese, South African and other radio astronomy institutes funded by their national research councils.
We thank the staff of JIVE, especially Bob Cambell, Zsolt Paragi, and Gabriele Surcis, for their assistance with the observations and correlation.
SDC thanks the STFC for support under a studentship.
We also thank the referee for providing many useful comments which greatly improved the paper, and Ian McHardy for useful discussions.
The research leading to these results has received funding from the European Commission Seventh Framework Programme (FP/2007-2013) under grant agreement No. 283393 (RadioNet3).
e-MERLIN is a National Facility operated by the University of Manchester at Jodrell Bank Observatory on behalf of STFC.
This research has made use of v0.7 of the e-MERLIN pipeline (DOI:10.5281/zenodo.10163).
The scientific results reported in this article are based in part on data obtained from the Chandra Data Archive.
This research has made use of the NASA/IPAC extragalactic database (NED) which is operated by the Jet Propulsion Laboratory, Caltech, under contract with the National Aeronautics and Space Administration.

\bibliographystyle{mn}
\bibliography{refs}

\begin{thebibliography}{43}
\expandafter\ifx\csname natexlab\endcsname\relax\def\natexlab#1{#1}\fi

\bibitem[{{Argo}(2015)}]{pipeline}
{Argo} M.~K., 2015, J. of Open Res. Software, 3(1):e2

\bibitem[{{Baan} {et~al.}(1986){Baan}, {Guesten}, \&
  {Haschick}}]{1986ApJ...305..830B}
{Baan} W.~A., {Guesten} R., {Haschick} A.~D., 1986, \apj, 305, 830

\bibitem[{{Baan} {et~al.}(1992){Baan}, {Rhoads}, \&
  {Haschick}}]{1992ApJ...401..508B}
{Baan} W.~A., {Rhoads} J., {Haschick} A.~D., 1992, \apj, 401, 508

\bibitem[{{Baars} {et~al.}(1977){Baars}, {Genzel}, {Pauliny-Toth}, \&
  {Witzel}}]{1977A&A....61...99B}
{Baars} J.~W.~M., {Genzel} R., {Pauliny-Toth} I.~I.~K., {Witzel} A., 1977,
  \aap, 61, 99

\bibitem[{{Bartel} {et~al.}(2002){Bartel}, {Bietenholz}, {Rupen}, {Beasley},
  {Graham}, {Altunin}, {Venturi}, {Umana}, {Cannon}, \&
  {Conway}}]{2002ApJ...581..404B}
{Bartel} N., {Bietenholz} M.~F., {Rupen} M.~P., {Beasley} A.~J., {Graham}
  D.~A., {Altunin} V.~I., {Venturi} T., {Umana} G., {Cannon} W.~H., {Conway}
  J.~E., 2002, \apj, 581, 404

\bibitem[{{Bietenholz} {et~al.}(2010{\natexlab{a}}){Bietenholz}, {Bartel}, \&
  {Rupen}}]{2010ApJ...712.1057B}
{Bietenholz} M.~F., {Bartel} N., {Rupen} M.~P., 2010{\natexlab{a}}, \apj, 712,
  1057

\bibitem[{{Bietenholz} {et~al.}(2010{\natexlab{b}}){Bietenholz}, {Soderberg},
  {Bartel}, {Ellingsen}, {Horiuchi}, {Phillips}, {Tzioumis}, {Wieringa}, \&
  {Chugai}}]{2010ApJ...725....4B}
{Bietenholz} M.~F., {Soderberg} A.~M., {Bartel} N., {Ellingsen} S.~P.,
  {Horiuchi} S., {Phillips} C.~J., {Tzioumis} A.~K., {Wieringa} M.~H., {Chugai}
  N.~N., 2010{\natexlab{b}}, \apj, 725, 4

\bibitem[{{Bloom} {et~al.}(2011){Bloom}, {Giannios}, {Metzger}, {Cenko},
  {Perley}, {Butler}, {Tanvir}, {Levan}, {O'Brien}, {Strubbe}, {De Colle},
  {Ramirez-Ruiz}, {Lee}, {Nayakshin}, {Quataert}, {King}, {Cucchiara},
  {Guillochon}, {Bower}, {Fruchter}, {Morgan}, \& {van der
  Horst}}]{2011Sci...333..203B}
{Bloom} J.~S., {Giannios} D., {Metzger} B.~D., {Cenko} S.~B., {Perley} D.~A.,
  {Butler} N.~R., {Tanvir} N.~R., {Levan} A.~J., {O'Brien} P.~T., {Strubbe}
  L.~E., {De Colle} F., {Ramirez-Ruiz} E., {Lee} W.~H., {Nayakshin} S.,
  {Quataert} E., {King} A.~R., {Cucchiara} A., {Guillochon} J., {Bower} G.~C.,
  {Fruchter} A.~S., {Morgan} A.~N., {van der Horst} A.~J., 2011, Science, 333,
  203

\bibitem[{{Carral} {et~al.}(1990){Carral}, {Turner}, \&
  {Ho}}]{1990ApJ...362..434C}
{Carral} P., {Turner} J.~L., {Ho} P.~T.~P., 1990, \apj, 362, 434

\bibitem[{{Combes} {et~al.}(1992){Combes}, {Braine}, {Casoli}, {Gerin}, \& {van
  Driel}}]{1992A&A...259L..65C}
{Combes} F., {Braine} J., {Casoli} F., {Gerin} M., {van Driel} W., 1992, \aap,
  259, L65

\bibitem[{{de Vries} {et~al.}(1997){de Vries}, {Barthel}, \&
  {O'Dea}}]{1997A&A...321..105D}
{de Vries} W.~H., {Barthel} P.~D., {O'Dea} C.~P., 1997, \aap, 321, 105

\bibitem[{{Dudik} {et~al.}(2005){Dudik}, {Satyapal}, {Gliozzi}, \&
  {Sambruna}}]{2005ApJ...620..113D}
{Dudik} R.~P., {Satyapal} S., {Gliozzi} M., {Sambruna} R.~M., 2005, \apj, 620,
  113

\bibitem[{{Fenech} {et~al.}(2010){Fenech}, {Beswick}, {Muxlow}, {Pedlar}, \&
  {Argo}}]{2010MNRAS.408..607F}
{Fenech} D., {Beswick} R., {Muxlow} T.~W.~B., {Pedlar} A., {Argo} M.~K., 2010,
  \mnras, 408, 607

\bibitem[{{Filho} {et~al.}(2002){Filho}, {Barthel}, \&
  {Ho}}]{2002ApJS..142..223F}
{Filho} M.~E., {Barthel} P.~D., {Ho} L.~C., 2002, \apjs, 142, 223

\bibitem[{{Fruscione} {et~al.}(2006){Fruscione}, {McDowell}, {Allen},
  {Brickhouse}, {Burke}, {Davis}, {Durham}, {Elvis}, {Galle}, {Harris},
  {Huenemoerder}, {Houck}, {Ishibashi}, {Karovska}, {Nicastro}, {Noble},
  {Nowak}, {Primini}, {Siemiginowska}, {Smith}, \&
  {Wise}}]{2006SPIE.6270E..60F}
{Fruscione} A., {McDowell} J.~C., {Allen} G.~E., {Brickhouse} N.~S., {Burke}
  D.~J., {Davis} J.~E., {Durham} N., {Elvis} M., {Galle} E.~C., {Harris} D.~E.,
  {Huenemoerder} D.~P., {Houck} J.~C., {Ishibashi} B., {Karovska} M.,
  {Nicastro} F., {Noble} M.~S., {Nowak} M.~A., {Primini} F.~A., {Siemiginowska}
  A., {Smith} R.~K., {Wise} M., 2006, in Society of Photo-Optical
  Instrumentation Engineers (SPIE) Conference Series, Vol. 6270, Society of
  Photo-Optical Instrumentation Engineers (SPIE) Conference Series

\bibitem[{{Gallimore} {et~al.}(1999){Gallimore}, {Baum}, {O'Dea}, {Pedlar}, \&
  {Brinks}}]{1999ApJ...524..684G}
{Gallimore} J.~F., {Baum} S.~A., {O'Dea} C.~P., {Pedlar} A., {Brinks} E., 1999,
  \apj, 524, 684

\bibitem[{{Giannios} \& {Metzger}(2011)}]{2011MNRAS.416.2102G}
{Giannios} D., {Metzger} B.~D., 2011, \mnras, 416, 2102

\bibitem[{{Guainazzi} \& {Bianchi}(2007)}]{2007MNRAS.374.1290G}
{Guainazzi} M., {Bianchi} S., 2007, \mnras, 374, 1290

\bibitem[{{Horiuchi} {et~al.}(2004){Horiuchi}, {Fomalont}, {Taylor}, {Scott},
  {Lovell}, {Moellenbrock}, {Dodson}, {Murata}, {Hirabayashi}, {Edwards},
  {Gurvits}, \& {Shen}}]{2004ApJ...616..110H}
{Horiuchi} S., {Fomalont} E.~B., {Taylor} W.~K., {Scott} A.~R., {Lovell}
  J.~E.~J., {Moellenbrock} G.~A., {Dodson} R., {Murata} Y., {Hirabayashi} H.,
  {Edwards} P.~G., {Gurvits} L.~I., {Shen} Z.-Q., 2004, \apj, 616, 110

\bibitem[{{Jones} {et~al.}(1981){Jones}, {Sramek}, \&
  {Terzian}}]{1981ApJ...246...28J}
{Jones} D.~L., {Sramek} R.~A., {Terzian} Y., 1981, \apj, 246, 28

\bibitem[{{Komossa}(2002)}]{2002RvMA...15...27K}
{Komossa} S., 2002, in Reviews in Modern Astronomy, Vol.~15, Reviews in Modern
  Astronomy, {Schielicke} R.~E., ed., p.~27

\bibitem[{{Komossa} \& {Dahlem}(2001)}]{2001astro.ph..6422K}
{Komossa} S., {Dahlem} M., 2001, ArXiv Astrophysics e-prints

\bibitem[{{Kulkarni} {et~al.}(1998){Kulkarni}, {Frail}, {Wieringa}, {Ekers},
  {Sadler}, {Wark}, {Higdon}, {Phinney}, \& {Bloom}}]{1998Natur.395..663K}
{Kulkarni} S.~R., {Frail} D.~A., {Wieringa} M.~H., {Ekers} R.~D., {Sadler}
  E.~M., {Wark} R.~M., {Higdon} J.~L., {Phinney} E.~S., {Bloom} J.~S., 1998,
  \nat, 395, 663

\bibitem[{{Lonsdale} {et~al.}(2006){Lonsdale}, {Diamond}, {Thrall}, {Smith}, \&
  {Lonsdale}}]{2006ApJ...647..185L}
{Lonsdale} C.~J., {Diamond} P.~J., {Thrall} H., {Smith} H.~E., {Lonsdale}
  C.~J., 2006, \apj, 647, 185

\bibitem[{{Lu} {et~al.}(1993){Lu}, {Hoffman}, {Groff}, {Roos}, \&
  {Lamphier}}]{1993ApJS...88..383L}
{Lu} N.~Y., {Hoffman} G.~L., {Groff} T., {Roos} T., {Lamphier} C., 1993, \apjs,
  88, 383

\bibitem[{{Mangum} {et~al.}(2013{\natexlab{a}}){Mangum}, {Darling}, {Henkel},
  \& {Menten}}]{2013ApJ...766..108M}
{Mangum} J.~G., {Darling} J., {Henkel} C., {Menten} K.~M., 2013{\natexlab{a}},
  \apj, 766, 108

\bibitem[{{Mangum} {et~al.}(2013{\natexlab{b}}){Mangum}, {Darling}, {Henkel},
  {Menten}, {MacGregor}, {Svoboda}, \& {Schinnerer}}]{2013ApJ...779...33M}
{Mangum} J.~G., {Darling} J., {Henkel} C., {Menten} K.~M., {MacGregor} M.,
  {Svoboda} B.~E., {Schinnerer} E., 2013{\natexlab{b}}, \apj, 779, 33

\bibitem[{{Marscher}(1983)}]{1983ApJ...264..296M}
{Marscher} A.~P., 1983, \apj, 264, 296

\bibitem[{{Mart{\'{\i}}-Vidal} {et~al.}(2011){Mart{\'{\i}}-Vidal}, {Tudose},
  {Paragi}, {Yang}, {Marcaide}, {Guirado}, {Ros}, {Alberdi},
  {P{\'e}rez-Torres}, {Argo}, {van der Horst}, {Garrett}, {Stockdale}, \&
  {Weiler}}]{2011A&A...535L..10M}
{Mart{\'{\i}}-Vidal} I., {Tudose} V., {Paragi} Z., {Yang} J., {Marcaide} J.~M.,
  {Guirado} J.~C., {Ros} E., {Alberdi} A., {P{\'e}rez-Torres} M.~A., {Argo}
  M.~K., {van der Horst} A.~J., {Garrett} M.~A., {Stockdale} C.~J., {Weiler}
  K.~W., 2011, \aap, 535, L10

\bibitem[{{Merloni} {et~al.}(2003){Merloni}, {Heinz}, \& {di
  Matteo}}]{2003MNRAS.345.1057M}
{Merloni} A., {Heinz} S., {di Matteo} T., 2003, \mnras, 345, 1057

\bibitem[{{Mezcua} \& {Prieto}(2014)}]{2014ApJ...787...62M}
{Mezcua} M., {Prieto} M.~A., 2014, \apj, 787, 62

\bibitem[{{Minchin} {et~al.}(2013){Minchin}, {Ghosh}, {Momjian}, \&
  {Salter}}]{2013AAS...22115706M}
{Minchin} R.~F., {Ghosh} T., {Momjian} E., {Salter} C.~J., 2013, in American
  Astronomical Society Meeting Abstracts, Vol. 221, American Astronomical
  Society Meeting Abstracts, p. 157.06

\bibitem[{{Muxlow} {et~al.}(2005){Muxlow}, {Pedlar}, {Beswick}, {Argo},
  {O'Brien}, {Fenech}, \& {Trotman}}]{2005MmSAI..76..586M}
{Muxlow} T.~W.~B., {Pedlar} A., {Beswick} R.~J., {Argo} M.~K., {O'Brien} T.~J.,
  {Fenech} D., {Trotman} W., 2005, \memsai, 76, 586

\bibitem[{{Nagar} {et~al.}(2005){Nagar}, {Falcke}, \&
  {Wilson}}]{2005A&A...435..521N}
{Nagar} N.~M., {Falcke} H., {Wilson} A.~S., 2005, \aap, 435, 521

\bibitem[{{Nagar} {et~al.}(2002){Nagar}, {Falcke}, {Wilson}, \&
  {Ulvestad}}]{2002A&A...392...53N}
{Nagar} N.~M., {Falcke} H., {Wilson} A.~S., {Ulvestad} J.~S., 2002, \aap, 392,
  53

\bibitem[{{O'Dea}(1998)}]{1998PASP..110..493O}
{O'Dea} C.~P., 1998, \pasp, 110, 493

\bibitem[{{Panessa} \& {Giroletti}(2013)}]{2013MNRAS.432.1138P}
{Panessa} F., {Giroletti} M., 2013, \mnras, 432, 1138

\bibitem[{{Phookun} {et~al.}(1998){Phookun}, {Anantharamaiah}, \&
  {Goss}}]{1998MNRAS.295..156P}
{Phookun} B., {Anantharamaiah} K.~R., {Goss} W.~M., 1998, \mnras, 295, 156

\bibitem[{{Satyapal} {et~al.}(2004){Satyapal}, {Sambruna}, \&
  {Dudik}}]{2004A&A...414..825S}
{Satyapal} S., {Sambruna} R.~M., {Dudik} R.~P., 2004, \aap, 414, 825

\bibitem[{{Springob} {et~al.}(2009){Springob}, {Masters}, {Haynes},
  {Giovanelli}, \& {Marinoni}}]{2009ApJS..182..474S}
{Springob} C.~M., {Masters} K.~L., {Haynes} M.~P., {Giovanelli} R., {Marinoni}
  C., 2009, \apjs, 182, 474

\bibitem[{{Strickland} {et~al.}(1997){Strickland}, {Ponman}, \&
  {Stevens}}]{1997A&A...320..378S}
{Strickland} D.~K., {Ponman} T.~J., {Stevens} I.~R., 1997, \aap, 320, 378

\bibitem[{{Terashima} \& {Wilson}(2003)}]{2003ApJ...583..145T}
{Terashima} Y., {Wilson} A.~S., 2003, \apj, 583, 145

\bibitem[{{van Driel} {et~al.}(1995){van Driel}, {Combes}, {Casoli}, {Gerin},
  {Nakai}, {Miyaji}, {Hamabe}, {Sofue}, {Ichikawa}, {Yoshida}, {Kobayashi},
  {Geng}, {Minezaki}, {Arimoto}, {Kodama}, {Goudfrooij}, {Mulder}, {Wakamatsu},
  \& {Yanagisawa}}]{1995AJ....109..942V}
{van Driel} W., {Combes} F., {Casoli} F., {Gerin} M., {Nakai} N., {Miyaji} T.,
  {Hamabe} M., {Sofue} Y., {Ichikawa} T., {Yoshida} S., {Kobayashi} Y., {Geng}
  F., {Minezaki} T., {Arimoto} N., {Kodama} T., {Goudfrooij} P., {Mulder}
  P.~S., {Wakamatsu} K., {Yanagisawa} K., 1995, \aj, 109, 942

\end{thebibliography}

\end{document}